\documentclass[aps,pra,reprint,showpacs,groupedaddress]{revtex4-1}

\usepackage[utf8]{inputenc}
\usepackage{amsmath}
\usepackage{bm}
\usepackage{graphicx}
\usepackage{siunitx}
\usepackage{hyperref}

\renewcommand{\cos}[1]{\textrm{cos}\left[#1\right]}
\renewcommand{\sin}[1]{\textrm{sin}\left[#1\right]}

\begin{document}
\title{Imaging the collective excitations of an ultracold gas using statistical correlations}
\author{Romain Dubessy, Camilla De Rossi, Thomas Badr, Laurent Longchambon and H\'el\`ene Perrin}
\address{Laboratoire de physique des lasers, CNRS, Universit\'e Paris 13, Sorbonne Paris Cit\'e, F-93430, Villetaneuse, France.}

\begin{abstract}
Advanced data analysis techniques have proved to be crucial for extracting information from noisy images. Here we show that principal component analysis can be successfully applied to ultracold gases to unveil their collective excitations. By analyzing the correlations in a series of images we are able to identify the collective modes which are excited, determine their population, image their eigenfunction, and measure their frequency. Our method allows to discriminate the relevant modes from other noise components and is robust with respect to the data sampling procedure. It can be extended to other dynamical systems including cavity polariton quantum gases or trapped ions.
\end{abstract}
\maketitle

\section{Introduction}
In the past few years, the degree of control of cold atom experiments has increased to an impressive level, from the control of the atomic interactions \cite{Chin2010} and the trapping geometry \cite{Gaunt2013}, to the creation and observation of many-body correlated systems \cite{Bloch2012} or the control at the single atom level \cite{Weitenberg2011}. In order to extract quantitative measurements from such experiments one has to analyze a large number of images \cite{Ketterle1999}, which are fitted and compared to theoretical models \cite{Dalfovo1999}. For instance mean-field models describe remarkably well quantum gases at low temperature, including their dynamics \cite{Castin1996,Kagan1996,Ketterle1999}. However these simple models are far from exploiting all the information contained in the images.

This has motivated the development of alternative \emph{model-free} approaches to analyze the experimental data. For example, with the minimal assumption that the image represents accurately the gas density profile, one can directly compute averaged observables to reveal the gas collective dynamics \cite{Fang2014}. It is also quite efficient to represent the signal in the frequency domain, using Fourier transforms, to isolate the system response to a resonant excitation \cite{Tey2013,Guajardo2013,Baur2013}. In some situations the noise itself contains a lot of information on the system \cite{Altman2004} that can be recovered by studying the correlations within the images \cite{Folling2005,Jacqmin2011}.

Here we show that a generic method of signal analysis, Principal Component Analysis (PCA) \cite{Jolliffe2002}, provides a unique tool to extract all the relevant information from cold atom absorption images, without having to rely on a specific model. This tool has already been used to perform filtering \cite{Segal2010,Chiow2011,Desbuquois2013}, extract the phase in an interferometric signal \cite{Sugarbaker2013,Dickerson2013} and identify the main noise sources in an experiment \cite{Farkas2014}. Recently it has been shown that PCA can be of interest to perform quantum state tomography \cite{Lloyd2014}. As part of multivariate signal analysis methods PCA is widely used in numerous applications dealing with large amounts of data \cite{Jolliffe2002}, to extract signals from a noisy background.

The main result of this paper is that PCA can be extended to the study of the elementary excitations of an ultracold atomic gas and allow the direct observation of the system normal modes. Normal modes or Bogoliubov modes of ultracold atomic gases are the elementary low energy excitations of the system \cite{Stringari1996,Dalfovo1999,Mewes1996}. They provide a unique insight into the system properties. For example they can reveal the collective superfluid behaviour of Bose \cite{Marago2000} and Fermi \cite{Kinast2004,Bartenstein2004,Nascimbene2009} gases or probe the system dimensionality \cite{Fang2014,Merloti2013b}. Recently an analysis of a set of absorption images using time to frequency domain transformation \cite{Tey2013,Guajardo2013,Baur2013} has been used to isolate a few low energy collective modes and study their damping. Having access to a method for data analysis which extracts the maximum information will be highly relevant for these studies.

This paper is organized as follows: the PCA method for noise filtering is discussed in section~\ref{sec:model}. We then show in section~\ref{sec:study} that the PCA enables a precise identification of the system low energy excitations. To support our analysis of the experimental data we compare our findings to the results of numerical simulations in section~\ref{sec:validation}. Finally we discuss in section~\ref{sec:discuss} the requirements for applying PCA to cold atom experiments and the possible improvements that may be achieved.

\section{Principal Component Analysis}
\label{sec:model}
Let us briefly recall how PCA proceeds \cite{Jolliffe2002}. More detail (including formulas) is given in \ref{sec:PCA}. We start from a particular data set, which in our case is an ensemble of absorption images where the signal is proportional to the integrated atomic density. We first compute the average of the data set and subtract this mean image from all the images, thus obtaining an ensemble of centered images. We then compute the covariance matrix of this ensemble. The diagonal elements of the covariance matrix contain the variance of the pixels and off-diagonal elements quantify correlations between pixels. By diagonalizing this matrix we recover the eigenvectors, called Principal Components (PCs), which are thus uncorrelated. This statistical independence ensures that uncorrelated noise sources are associated to different principal components \cite{Jolliffe2002}.

Our experiment is described in detail in reference \cite{Merloti2013a}. Briefly, we produce a quantum degenerate gas of $^{87}$Rb atoms confined in a radio-frequency (rf) dressed magnetic quadrupole trap. We can dynamically control the precise trap shape by varying the magnetic or rf fields, which results in selective excitations of the gas normal modes \cite{Merloti2013a,Merloti2013b}. We measure the gas properties by performing in-situ absorption imaging along the strongly trapped vertical direction. The peak optical density is kept below 6 by repumping only a small fraction of the cloud from the $F=1$ hyperfine ground state to the cycling transition. We carefully calibrate the imaging system following reference \cite{Reinaudi2007}. The gases we consider in this paper are in the quasi two-dimensional regime: the excitations along the imaging axis are frozen and the dynamics occurs only in the horizontal plane. In this plane the system is well described by a harmonic oscillator \cite{Merloti2013a}. We apply the PCA to the study of the mode dynamics in an anisotropic quasi two-dimensional gas ($\mu/(\hbar\omega_z)\sim1.5$) with $\omega_x=2\pi\times\SI{33}{Hz}$ and $\omega_y=2\pi\times\SI{44}{Hz}$.

\begin{figure}[h]
\begin{center}
\includegraphics[width=8cm]{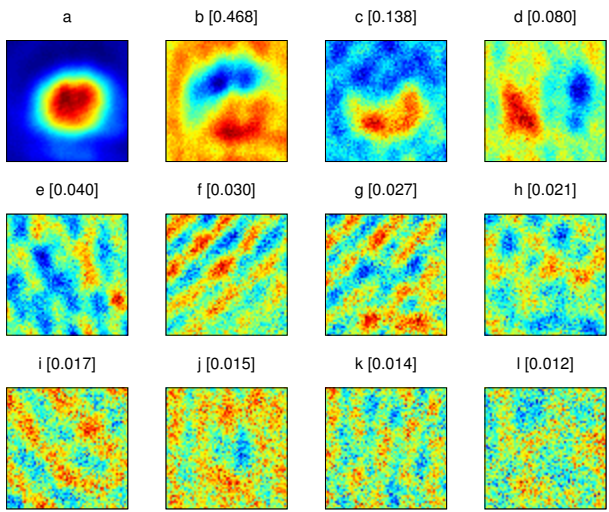}
\caption{\label{fig:expnoise}(Color online)
Noise analysis using PCA. Figure~a: averaged image ($61\times 61$ pixels), and figures~b-l: the eleven largest principal components, sorted by decreasing eigenvalue. The number between square bracket is the eigenvalue of the principal component, expressed as a fraction of the total variance. The color scale is arbitrary for each image. The field of view is $61\times\SI{61}{\micro\meter^2}$.
}
\end{center}
\end{figure}

As an example of application, figure~\ref{fig:expnoise} displays the outcome of PCA applied to 27 images acquired in the same conditions. Due to variations of the stray light during image acquisition, fluctuations of atom number in the experiment or mechanical vibrations, the images are not exactly identical. The PCA decomposition identifies all these sources of noise, and we can identify them with a principal component: figure~\ref{fig:expnoise}c probably accounts for atom number fluctuations while figure~\ref{fig:expnoise}b and d indicates a small jitter of the camera position \cite{Segal2010,Farkas2014}. Higher order components, see figure~\ref{fig:expnoise}e to h, reflect the presence of diffraction fringes on the probe beam intensity profile. For each of these components, the corresponding eigenvalue accounts for the fraction of the total variance due to the associated noise source.

Conversely, when the data set results from the variation of a parameter, PCA allows to probe the sensitivity of the system to this parameter. In particular, if the system is initially excited and evolves, measurements taken at different times allow to recover this variation as a principal component. In this paper, we exploit this possibility to directly measure the normal modes of a quantum degenerate gas confined in a harmonic trap.

\section{Evidencing the excited modes}
\label{sec:study}
\begin{figure}[h]
\begin{center}
\includegraphics[width=8cm]{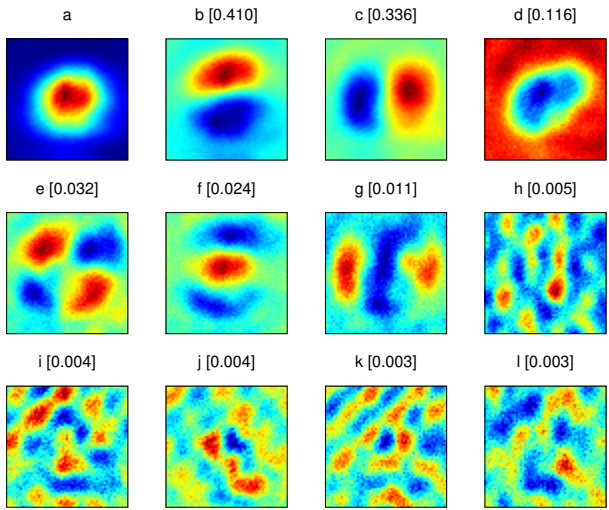}
\caption{\label{fig:scissormodeExp}(Color online)
First principal components of an ensemble of 133 images ($61\times61$ pixels) sampling a time interval of $\SI{100}{ms}$. Figure~a is the mean image of the data set, containing the averaged density profile and the subsequent images (b to l) are the first principal components (sorted by decreasing eigenvalue). The number between square brackets is the corresponding eigenvalue, expressed as a fraction of the total variance. The color scale is arbitrary for each image.}
\end{center}
\end{figure}

We make use of our highly versatile trap potential to excite simultaneously several low energy eigenmodes of the gas. Namely we displace the trap minimum, we rotate the trap axes and change slightly the trap frequencies. In the new trap the gas is strongly out of equilibrium and we record its evolution by taking images for different holding times in the trap, covering a time span of $\SI{100}{ms}$. Figure~\ref{fig:scissormodeExp} shows the result of PCA for this data set (133 images). Compared to figure~\ref{fig:expnoise} we see that the principal components have changed.

Let us now identify the first principal components. The first two PCs (see figure~\ref{fig:scissormodeExp}b and c) display a two-lobe pattern oriented respectively along the columns and the rows of the images: this is characteristic of a dipole oscillation of the cloud. This center of mass motion is due to the trap minimum displacement during the excitation process. The third PC (see figure~\ref{fig:scissormodeExp}d) indicates a global variation of the signal over the whole cloud, which can be interpreted as atom number fluctuations in the experiment. Part of these fluctuations are due to the fact that the lifetime in the trap is limited and atoms are lost as the holding time increases\footnote{This PC displays a tilted shape, compared to the mean density profile. We attribute this to the fact that atom number fluctuations in this experiment were dominated by fluctuations in the repumping beam that comes from the side of the cloud.}. The fourth PC (see figure~\ref{fig:scissormodeExp}e) possesses a striking spatial pattern with four lobes characteristic of a scissors excitation \cite{DGO1999}. Note that the lobes are oriented at $45^\circ$ with respect to the trap axes (aligned with the first two PCs) as expected. The next two PCs (see figure~\ref{fig:scissormodeExp}f and g) look like compression modes of the gas, with a density depletion at the center of the cloud and a correlated augmentation of the density on the sides of the cloud.

The PCs are presented by decreasing eigenvalue, meaning that they account for less and less variations in the original data set. For this particular experiment the center of mass oscillation is the dominant excitation in the cloud, followed by the response to the rotation of the trap axis and marginally by the compression of the trap. In another experiment (not shown) where the trap rotation was not performed we have verified that no PC displayed the spatial pattern of a scissors excitation.

\begin{figure}[h]
\begin{center}
\includegraphics[width=8cm]{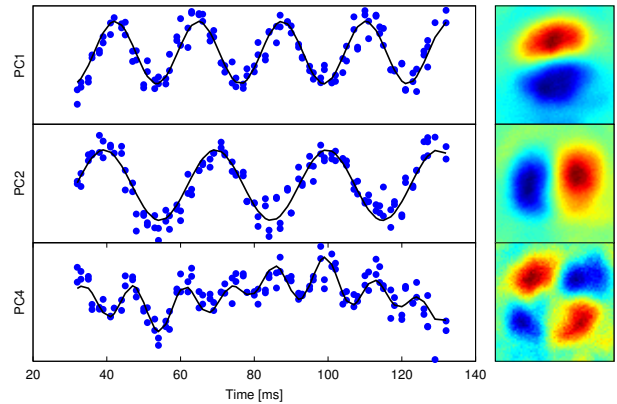}
\caption{\label{fig:scissorTime}(Color online)
Filled blue circles: time dependent weight of the two dipoles and the scissors components. Black solid line: sinusoidal fit to the data. The vertical scale is arbitrary and independent for each curve. The first principal component can be identified to the strongest horizontal harmonic trap direction, oscillating at $\SI{44}{Hz}$, the second to the weakest corresponding to a frequency of $\SI{33}{Hz}$. The third component exhibits a more complicated behavior with oscillations at $\SI{12}{Hz}$, $\SI{55}{Hz}$, and $\SI{77}{Hz}$. We estimate at the $\SI{1}{Hz}$ level the uncertainty on the frequency determination by the fitting procedure.
}
\end{center}
\end{figure}

This analysis of the PCs is supported by the study of the time dependent oscillations of the associated weights, computed by projecting the centered original data set on the PCs. The result of this computation is displayed on figure~\ref{fig:scissorTime}, for the dipoles and scissors components. Let us focus first on the first two weights: they exhibit sinusoidal oscillations at the expected trap frequencies ($\SI{44}{Hz}$ and $\SI{33}{Hz}$). This supports the fact that PCA has correctly identified as independent components the center of mass motion of the cloud along the trap axes. The scissors component displays a more complex oscillation pattern. We find that the best fit to the data is given by a sum of three sinusoids, at frequencies $\SI{12}{Hz}$, $\SI{55}{Hz}$ and $\SI{77}{Hz}$. This is related to the fact that the scissors component found by PCA is sensitive both to the collective response of the superfluid part and to the collisionless oscillations of the normal part of the gas \cite{DGO1999}. The simultaneous presence of these three frequencies has been evidenced in a three dimensional Bose-Einstein condensate \cite{Marago2001} where simultaneous measurement of the superfluid and normal part of the cloud rotations were obtained by a bimodal distribution fit to the density profiles. Here we note that the same PC gives access to both the superfluid and the normal response to the rotation of the trap axes which might be used to measure their relative amplitudes.

Let us stress that the PCA is able to separate the contributions of the different modes in a given experiment which could help to design better excitation pattern or focus on higher order modes \cite{Guajardo2013}. In particular, being able to measure on the same data set the dipole mode frequencies gives access to the natural system clock \cite{Stringari1996}. Therefore PCA gives access to direct comparison between measured frequencies and predictions. Moreover, for the data set used in figure~\ref{fig:scissormodeExp}, we find that the simple hydrodynamic models of references \cite{Castin1996,Kagan1996} fail to extract these frequencies present in the oscillation of the density, probably due to the fact that several collective modes are simultaneously excited. In this case it is really essential to use a \emph{model-free} approach to analyze the data set.

\section{Comparison to numerical simulations}
\label{sec:validation}
We pursue our investigation numerically in order to compare the principal components to normal modes. We use a zero temperature two-dimensional mean field model of our cloud and perform a numerical time-dependent simulation which mimics the experimental sequence. We then extract the simulated density profiles using a regular time sampling, thus obtaining a data set of 152 computed images. We finally compare the PCA of this data set to the actual normal modes of the trap, computed using the Bogoliubov-de Gennes equations. Details on the simulations are given in \ref{sec:simulations}.

\begin{figure*}[t!]
\begin{center}
\begin{minipage}[c]{0.47\linewidth}
\includegraphics[width=\linewidth]{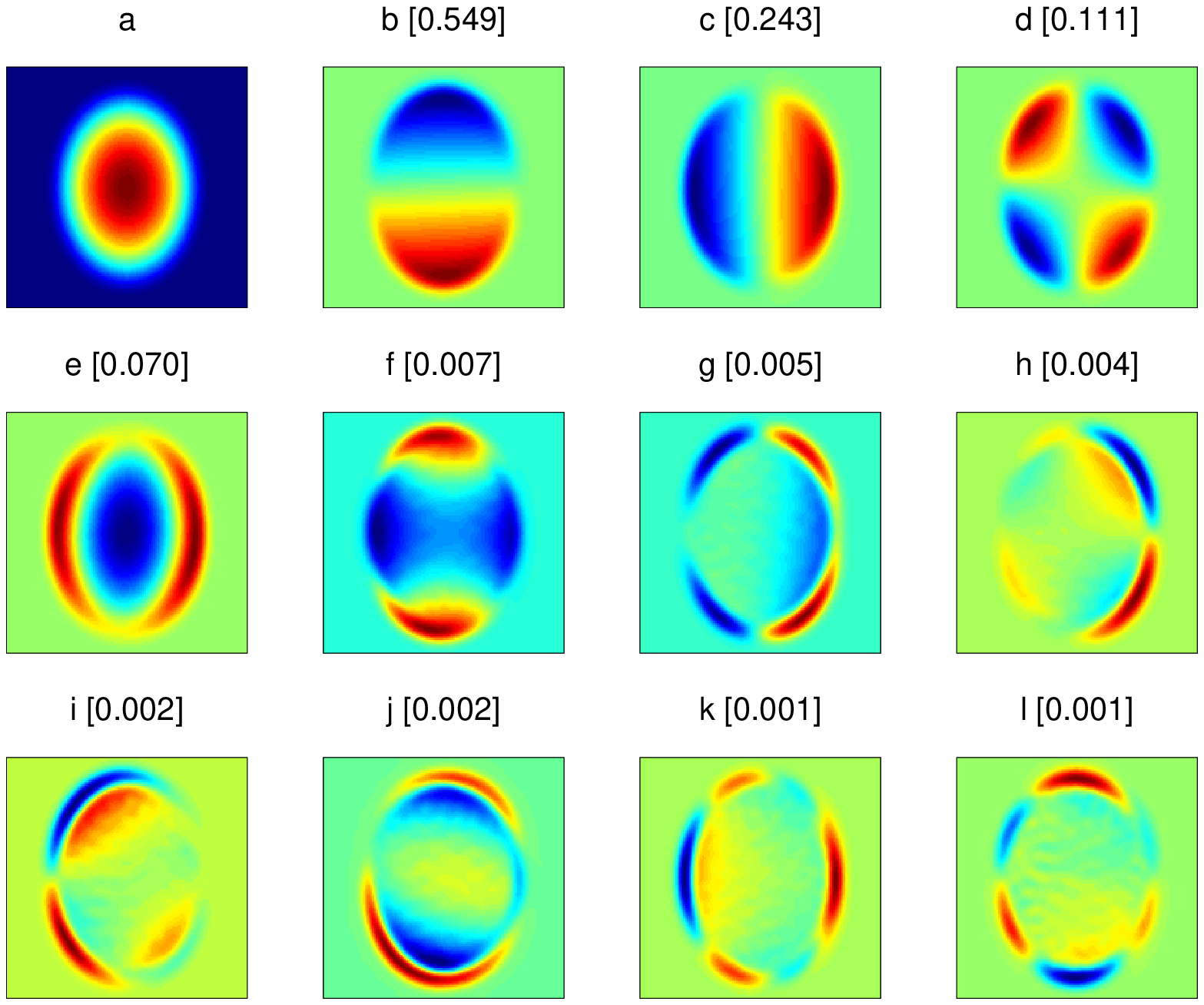}
\end{minipage}\hfill
\begin{minipage}[c]{0.47\linewidth}
\includegraphics[width=\linewidth]{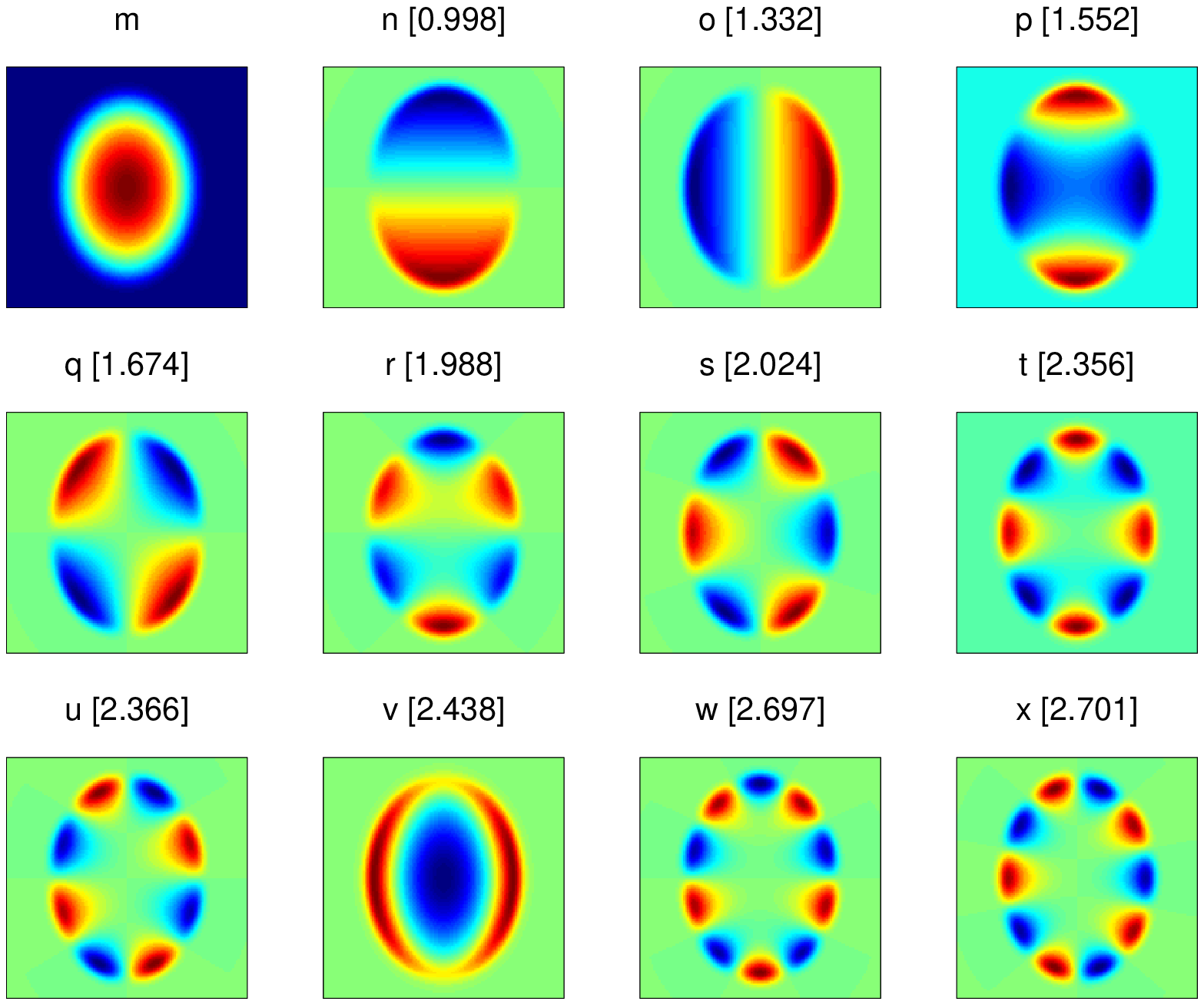}
\end{minipage}
\caption{\label{fig:scissormodeTheory}(Color online) Comparison of the principal components and the exact normal modes of the trapped cloud. Left panel: principal component analysis of the cloud shape during oscillations for six trap periods (of the weakest axis). The average cloud (image a) and the first eleven principal components are shown by decreasing eigenvalue, indicated between square brackets (and normalized to the total variance). Right panel: density profile of the cloud (image m) and the first eleven Bogoliubov modes for a gas at rest in the final trap. The modes are sorted by increasing mode frequency, indicated between square brackets in units of $\omega_x$.}
\end{center}
\end{figure*}
Figure~\ref{fig:scissormodeTheory} displays the result of the simulations. Let us first focus on the output of PCA (left panel): the first few PCs are like those of figure~\ref{fig:scissormodeExp}, except for the atom number fluctuations which are not taken into account in the simulation. In particular, dipole, scissors, monopole-like and quadrupole-like patterns are present (see respectively figure~\ref{fig:scissormodeTheory}b to f).

This interpretation is supported by the display of the normal modes (right panel), and in particular by the density profiles of figures~\ref{fig:scissormodeTheory}n-q and v. To compare these profiles quantitatively, we compute the scalar product between the principal component and the eigenmode images. We find a high degree of overlap for the largest five principal components (dipoles: $99.7\%$ and $99.4\%$, scissors: $98.5\%$, monopole-like $98.8\%$ and quadrupole-like $89.2\%$) when projected onto the corresponding eigenmode. This supports the idea that the largest principal components can indeed be identified with a well defined normal mode (see also \ref{sec:identification}).

\begin{table}
\begin{tabular}{lccc}
\hline
label & $\omega_{\rm pca}$ & $\omega_{\rm diag}$ & $\omega_{\rm th}$\\\hline
dipole (x) & 0.999 & 0.998 & 1\\
dipole (y) & 1.332 & 1.332 & 1.334\\
quadrupole & 1.547 &  1.552 & 1.548\\
scissors & 1.674 &  1.674 &  1.667\\
monopole & 2.441 &  2.438 & 2.438
\end{tabular}
\caption{\label{tab:Theory}
Comparison of the first principal components oscillation frequency $\omega_{\rm pca}$ with the Bogoliubov modes frequency $\omega_{\rm diag}$ and a hydrodynamic Thomas-Fermi model $\omega_{\rm th}$. All frequencies are given in units of the smallest dipole frequency $\omega_x$.
}
\end{table}

To confirm this result, we compare the oscillation frequency of the principal components (obtained by fitting a sinusoidal function to the time dependent weight of the simulated density profiles) $\omega_{\rm pca}$ to the frequency of the mode given by the Bogoliubov-de Gennes equations $\omega_{\rm diag}$ and to an analytic hydrodynamical model $\omega_{\rm th}$ \cite{Stringari1996}. The results obtained with the data of figure~\ref{fig:scissormodeTheory} are reported in table~\ref{tab:Theory}. We find that for the largest principal components the simple sinusoidal behavior correctly fits the data and gives a value compatible with the Bogoliubov-de Gennes theory, within the numerical uncertainty\footnote{The accuracy of $\omega_{\rm pca}$ determination is limited by the total simulation time to the $5\times10^{-3}\omega_x$ level, allowing to resolve the small $\sim0.1\omega_x$ frequency difference between the scissors and the quadrupole components. We evaluate the order of magnitude of the spatial discretization numerical error by comparing the computed dipoles frequencies $\omega_{\rm diag}$ to their exact theoretical value $\omega_{\rm th}$. This effect is at the level of $\sim2\times10^{-3}\omega_x$.}.

For the collective modes, we expect the correct value for the mode frequency to be given by the diagonalization procedure, as the hydrodynamical model is only approximate. There is an excellent agreement between $\omega_{\rm pca}$ and $\omega_{\rm diag}$ for the values reported in table~\ref{tab:Theory}, thus validating our experimental findings. However this is not true for the principal components with a small variance, which exhibit complex temporal behaviors. We observe that these components do not have a significant overlap with one of the modes: PCA is not able to identify them.

We conclude from this example that PCA provides a robust way of evidencing the dominantly excited modes in an out of equilibrium ultracold gas. Once the relevant components are isolated, PCA allows to extract the mode time dependence, without having to rely on a model for the atomic response.

\section{Discussion}
\label{sec:discuss}
We now discuss the requirements for PCA to be efficient and compare it with Fourier analysis. PCA is a statistical method: the data set has to span a sufficiently large number of configurations for the correlations between two different normal modes to average to zero. In particular, to resolve two different modes with close frequencies, the total acquisition time has to be larger than the beat note period. However it is not necessary to use an even sampling during this time period. In addition if the populations in the two excitations are very different, resulting in very different contributions to the variance, PCA separates them efficiently, even for shorter observation times, see the discussion in \ref{sec:identification}.

Fourier transformation methods can also be quite efficient for identifying collective modes \cite{Tey2013,Guajardo2013,Baur2013}. However they come with stronger constraints: the time sampling must be regular and the total time should be a multiple of the signal time period. It supposes an \emph{a priori} knowledge of the signal frequency which may have to be determined iteratively. Moreover Fourier transform gives only access to frequencies which are multiple of a fundamental frequency, which complicates the analysis for systems with multiple excitations (see supplementary data). Finally, a white noise contributes to each Fourier component, whereas it is naturally filtered out in PCA. PCA is not subject to such constraints: if we reduce the size of the data set used in section~\ref{sec:validation}, for example by keeping only one image out of ten, PCA is still able to find out PCs close to the excited eigenmodes (dipoles, scissors and monopole-like with $95\%$ fidelity, but the quadrupole-like component is absent, see supplementary data), even if the Nyquist-Shannon sampling theorem is not verified any longer\footnote{However the Nyquist-Shannon theorem is not violated, as for such a low sampling no information on the modes frequencies can be obtained from the time dependent weights.}. In that sense PCA is more efficient than Fourier methods.

In conclusion we have shown that, beyond noise filtering  \cite{Segal2010,Chiow2011,Desbuquois2013}, PCA provides a powerful statistical tool to analyze experimental as well as numerical data sets. When applied to time-dependent systems, it allows for a \emph{model-free} discrimination of the normal modes and to the measurement of their populations and frequencies. We expect PCA to be particularly relevant for the study of samples where fluctuations play a major role in the physics. Examples include the random creation of defects in the Kibble Zurek mechanism \cite{Corman2014} or the correlations between vortices and anti-vortices in a two dimensional superfluid \cite{Choi2013}. We note that PCA is a very general method and would be suitable for other systems with time dependent signals. In cavity polariton quantum gases, where images can be taken in real time, PCA will allow to extract the relevant information inside a large data set \cite{Amo2011}. Finally, cold trapped ions systems behave as crystals supporting many collective modes, which could be studied using PCA \cite{Landa2012}.

We envision that PCA is suited to perform \emph{Bogoliubov spectroscopy}, in the spirit of the method used in references \cite{Tey2013,Guajardo2013}. A mode largely excited by a resonant excitation will be easily identified by PCA and its frequency precisely determined by measuring its eigenvalue with respect to the modulation frequency. In contrast to Fourier methods, PCA can identify the dominantly excited mode using samples covering only one oscillation period of the mode either by recording the time evolution or by varying the excitation phase. This property should prove useful in particular to study over-damped modes.

\begin{acknowledgments}
LPL is UMR 7538 of CNRS and Paris 13 University. We acknowledge helpful discussions with Aur\'elien Perrin.
\end{acknowledgments}

\appendix

\section{Principal Component Analysis}
\label{sec:PCA}
We provide here a short recipe to apply PCA to the analysis of density profiles. Other examples of applications are given in references \cite{Jolliffe2002,Segal2010}. We stress that the mathematical formalism is quite simple and that most data analysis softwares provide standard implementation of PCA. We are interested in density profile images and assume that the pixels of each image are stored (row wise) in a single vector. The first step is to center the data set by computing the average image and subtract it from each image. The whole data set can then be stored in an $N\times P$ matrix, denoted $B$, where $N$ is the number of images and $P$ is the number of pixels. Thus $B_{i,j}$ contains the $j$-th pixel of the centered image $i$.

Next we want to compute the eigenvalues of the covariance matrix $S=B^TB/(N-1)$ where $B^T$ is the transpose of matrix $B$. This $P\times P$ matrix is in general quite large so it is hard to diagonalize it directly. However it is quite simple to show that its rank is at most $N$. Indeed, assuming that $X$ is an eigenvector of $S$ with eigenvalue $\lambda$ (meaning $SX=\lambda X$), it is straightforward to verify that $Y=BX$ is an eigenvector of the square $N\times N$ symmetric matrix $\Sigma=BB^T/(N-1)$ with the same eigenvalue $\lambda$. Therefore $S$ and $\Sigma$ have the same spectrum, of at most $N$ real eigenvalues. Knowing an eigenvector $Y$ of $\Sigma$, the corresponding eigenvector of $S$ is simply $X=B^TY$. Finally let us stress that these vectors are orthogonal since the $S$ matrix is real and symmetric. We define the associated PCs by normalizing the eigenvectors to unity. In the case of both a large number of pixels and a large number of images the diagonalization of $S$ and $\Sigma$ is hard to compute. However since we are \emph{a priori} only interested in the PCs with the largest variance they can be efficiently computed by iterative methods \cite{Lehoucq1998}.

The PCs provide an orthonormal basis spanning the subspace of the data set and therefore each original image can be represented as a sum of the mean image and the weighted contributions from each principal component. These weights are obtained by projecting the centered image onto the corresponding principal component. By selecting only relevant principal components the noise can be partially filtered out of the reconstructed images \cite{Jolliffe2002}.

\section{Numerical simulations}
\label{sec:simulations}
We model our system by a zero temperature bi-dimensional Gross-Pitaevskii equation:
\begin{equation}
\imath\frac{\partial}{\partial t}\psi=\left(-\frac{\bm{\nabla}^2}{2}+V(x,y)+g_{2D}N\left|\psi\right|^2\right)\psi,
\label{eqn:gpe2D}
\end{equation}
where $t$ is expressed in units of $\omega_x^{-1}$, $x$ and $y$ in units of $a_x=\sqrt{\hbar/(M\omega_x)}$, and $\psi\equiv\psi(x,y,t)$ in units of $a_x^{-1}$. $M$ is the atomic mass, $N$ the number of atoms and $g_{2D}=\sqrt{8\pi}a/a_z$ is the reduced coupling constant, where $a$ is the contact interaction scattering length and $a_z=\sqrt{\hbar/(M\omega_z)}$ is the size of the vertical harmonic oscillator ground-state. The potential reads:
\begin{eqnarray}
&&V(x+x_0,y+y_0)=\\
&&\frac{\alpha}{2}\left[\left(x\cos{\theta}+y\sin{\theta}\right)^2+\epsilon\left(x\sin{\theta}-y\cos{\theta}\right)^2\right],\nonumber
\label{eqn:potential}
\end{eqnarray}
where $\epsilon=\omega_y^2/\omega_x^2$ quantifies the trap in plane anisotropy and the arbitrary angle $\theta$ allows to rotate the trap axes. The auxiliary parameters $x_0$, $y_0$ and $\alpha$ can be used to induce a trap displacement and compression. Table~\ref{tab:potential} details the value of the parameters appearing in \eqref{eqn:potential} before and after the excitation.

\begin{table}
\begin{tabular}{lccccc}
\hline
 & $\alpha$ & $\epsilon$ & $x_0$ & $y_0$ & $\theta$\\\hline
initial & 0.95 & 1.68 & 0.5 & 0.25 & 10$^\circ$\\
final & 1 & 1.78 & 0 & 0 & 0$^\circ$
\end{tabular}
\caption{\label{tab:potential}
Value of the trapping potential parameters used in the simulation before and after the excitation.}
\end{table}

The numerical wave function is represented on a square $128\times128$ grid with an equivalent full width of $15a_x$ in both $x$ and $y$ directions. For the computations we used $g_{2D}N=1000$, matched to the experimental conditions.

We use this model to compare the outcome of two numerical computations. On the one hand we mimic the experiment described in section~\ref{sec:study} by \emph{a}) computing the system ground state for the initial potential using an imaginary time evolution algorithm; \emph{b}) using this result as the input of a real time evolution in the final potential; and \emph{c}) performing PCA on regularly sampled density profiles during the evolution. The evolution algorithm relies on a time splitting spectral method, from $t=0$ to $t=37.7$ (in dimensionless units, corresponding to 6 periods of the weakest trap axis) using a time step of $10^{-3}$. The total time is chosen to be close to a multiple of both dipole modes oscillation period (6 periods and 8 periods respectively): this ensures that the average density profile computed in PCA is not skewed. The sampling is performed every $T_s=0.126$. The result of this procedure is shown in the left panel of figure~\ref{fig:scissormodeTheory}.

On the other hand we directly compute the small excitation spectrum using Bogoliubov-de Gennes equations obtained from the linearisation of \eqref{eqn:gpe2D} around the system ground state in the final trap. This implies the ability to diagonalize a square $2^{15}\times 2^{15}$ matrix which is quite challenging. Fortunately this matrix is sparse and we are interested only in the lowest energy part of the spectrum, which means we do not have to compute all the eigenstates. We designed a fast custom C program that uses a combination of an iterative method \cite{Lehoucq1998} with an efficient sparse matrix library \cite{Davis2004} to compute the relevant eigenvectors. The result of this procedure is displayed in the right panel of figure~\ref{fig:scissormodeTheory}.

\section{Identification of the principal components with the normal modes}
\label{sec:identification}
We have shown that PCA is very efficient to identify the normal modes of an excited ultracold gas. This may be surprising but can be understood in the framework of small excitations. Using a hydrodynamic model \cite{Stringari1996}, the gas out of equilibrium density profile may be expanded as: $\rho(\bm{r},t)=\rho_0(\bm{r})+\sum_k c_k\cos{\omega_k t+\phi_k}f_k(\bm{r})$, where $k$ labels the normal mode of frequency $\omega_k$, $f_k(\bm{r})$ describes the mode normalized density profile and $c_k$ is related to the mode population.

In the experiment we observe the gas only at discrete times $\{t_n\}_{n\in[1,N]}$ and positions $\{\bm{r}_i\}_{i\in[1,P]}$ and therefore we can write the $i$-th pixel of the $n$-th image as $\rho(\bm{r}_i,t_n)=\rho_0(\bm{r}_i)+\sum_k c_k\cos{\omega_k t_n+\phi_k}f_k(\bm{r}_i)+\varepsilon(\bm{r}_i,t_n)$, where we added a pixel and time dependent noise contribution $\varepsilon(\bm{r}_i,t_n)$. PCA starts with the evaluation of the centered data set, by averaging over the sampling times $\{t_n\}$: $B_{n,i}=\sum_k c_k\cos{\omega_k t_n+\phi_k}f_k(\bm{r}_i)+\varepsilon(\bm{r}_i,t_n)+\delta(\bm{r}_i)$, where the $\delta(\bm{r}_i)$ term is close to zero for a total sampling time $T\gg(\min_{k}\omega_k)^{-1}$.

Then the $S$ matrix elements can be written as:
\begin{equation}
S_{i,j}=
\frac{N}{2(N-1)}\sum_k c_k^2f_k(\bm{r}_i)f_k(\bm{r}_j)+\Delta(\bm{r}_i,\bm{r}_j),
\label{eqn:covariance}
\end{equation}
where the term $\Delta(\bm{r}_i,\bm{r}_j)$ is the effective noise covariance between pixels $i$ and $j$, due to the initial noise distribution and finite sampling size induced errors. Providing that the $\Delta(\bm{r}_i,\bm{r}_j)$ term is small enough it is straightforward to verify that the principal components of matrix $S$ are the vectors $\{f_k(\bm{r}_i)\}_i$, with eigenvalue $\sim c_k^2/2$. In particular this is true for $T\gg\left(\min_{k\neq k^\prime}|\omega_k-\omega_{k^\prime}|\right)^{-1}$. This constraint on $T$ is more stringent than the previous one, especially when two normal modes are close to degeneracy. However if these two modes have small populations they have a small contribution to the $\Delta(\bm{r}_i,\bm{r}_j)$ term.

The conclusion of this analysis is twofold. On the one hand PCA correctly identifies the most excited\footnote{Indeed if $c_k^2/2$ for the $k$-th mode is comparable to the average value of $\Delta(\bm{r}_i,\bm{r}_j)$, then this mode remains mixed with noise components.} eigenmodes of the system. On the other hand, the total time sampling should be large enough to resolve the beat note between these modes. For practical purposes we empirically found that taking $T$ equal to one beat note period is sufficient, see for example figure~\ref{fig:scissorTime}.

\end{document}